# Optical bistability in nonlinear system with two loops of feedback

[1]George P. Miroshnichenko, [2]Alexander I. Trifanov

Saint-Petersburg State University of Information Technologies, Mechanics and Optics
Kronverksky 49, St.-Petersburg, 197101, Russia
[1]gpmirosh@gmail.com, [2]alextrifanov@gmail.com

## ABSTRACT

A model of nonlinear optical system surrounded by two loops of feedback is investigated. The cell with the vapor of rubidium $\Lambda$ - type atoms is taken in the capacity of nonlinear element. Two modes of near-resonant electromagnetic field interacting with the cell are involved in the feedback. Two-dimensional optical bistability domain in location of input field intensities is obtained and dependence of its form and magnitude from the system parameters (photon detunings, feedback factor etc.) is investigated. "Input – output" relations corresponding to different trajectories in the bistability domain are obtained. Cross-hysteresis is studied.

Atom-field interaction, nonlinear interaction, feedback, optical bistability.

## INTRODUCTION

Recently optical and electro-optical devices with feedback control are used intensively in information technologies. Generating of laser pulses for transmitting through the optical waveguide and error correction during the information processing are remarkable instances here. In-depth information about feedback applications in classical optical processing can be found[1]. Quantum theory of feedback[2] and its applications in designing of quantum memories[3] and implementation of quantum processing with embedded control[4] are under consideration recently.

In nonlinear optical systems surrounded by loop of feedback optical bistability (OB) may appear[5]. Input – output relations are not unique in this case and determined by previous evolution of the system. Existence of several output signal levels corresponded to the same input allows implementation of optical triggers, transistors and other basic elements of information processing[1, 5]. The main reasons of OB are the saturation of atomic transition and nonlinear relationship between refractive index and electromagnetic field. The first of them associated with absorptive type of OB and second with dispersive type[5, 6, 7]. Two-state atom interacting with a mode of electromagnetic field in unidirectional ring cavity is the simplest model of optical system with OB[8, 9]. To control the hysteresis parameters (thresholds, etc.) effectively one can use three-state atom of $\Lambda$-type instead of two-state atom. This allows to insert an addition (controlling) electromagnetic field into the system[10,11]. Varying the parameters of controlling field gives the comprehensive facilities in the changing of hysteresis parameters. Optical multistability (OM) in such system was predicted[10]. There are several works[12, 13, 14] devoted to experimental observation of OB and OM by using the interaction between rubidium $\Lambda$-atoms and electromagnetic field. Particularly, the transition from instability operation mode to bistable regime under different system parameters is investigated[12]. The relation between quantity of OB thresholds and intensity of controlling field is studied[13]. OM is observed[14]. It appears then two type of OB are observed simultaneously. The requirement of OM observation is high optical density of the atomic vapors. Note, that nonlinear relation of absorption factor from the intensity (amplitude) of the field is not unique which causes hysteresis. It is possible to observe the hysteresis-type curve by considering the nonlinear relation between frequency detuning and absorption factor[15]. Here we study the case then both two modes of electromagnetic field interacting with $\Lambda$-atom is inserted in the feedback. Absorptive type of OB in semiclassical mean field approximation and in stationary limit is under consideration.

# A MODEL

Let us put the cell with the vapors of $\Lambda$-type atoms into the unidirectional cavity as it is pictured on fig.1. Reflection and transmission coefficients of semitransparent mirrors $M_j$, $j=1,2$ are labeled $R_j$ and $T_j$ correspondingly $(T_j + R_j = 1)$. For simplification we assume the remaining mirrors are of ideal reflectivity. Two modes $E_1$ and $E_2$ of classical electromagnetic field with the frequencies $\omega_1$ and $\omega_2$ correspondingly are excited in the cavity. Let us denote their input intensities by $I_j^0$ and the intensity at the outcome of the cell by $I_j^{out}$. A part of radiation transmitted through the cell being reflected from the mirrors is added to the input signal and come to input of the cell again. Neglecting the effects of interference between input and redirected field we write the expression for intensities at the input of the cell:

$$I_j^{in} = I_j^0 + R_j I_j^{out}. \tag{1}$$

Denoting the absorption factors of the cell by $\eta_j = \eta_j\left(I_1^{in}, I_2^{in}\right)$ we can write:

$$I_j^{out} = \eta_j\left(I_1^{in}, I_2^{in}\right) I_j^{in}. \tag{2}$$

Let us rewrite the expression (1) using (2). The following system of nonlinear equations describing the intensities evolution in the cavity is obtained:

$$\eta_j\left(I_1^{in}, I_2^{in}\right) = \frac{1}{R_j} - \frac{1}{R_j I_j^{in}} I_j^0. \tag{3}$$

To evaluate the coefficient of absorption $\eta_j\left(I_1^{in}, I_2^{in}\right)$ it is necessary to solve the master equation for the case of atom – field interaction (see fig. 2). Let us consider the modes $E_1$ and $E_2$ interacting with atomic transitions $|1\rangle \to |2\rangle$ and $|3\rangle \to |2\rangle$ correspondingly. Hamiltonian operator for the system in the rotating frame and resonant approximation may be written as follows:

$$H = \Delta_1 \sigma_{22} + \Delta_2 \sigma_{33} + \Omega_1(\sigma_{12} + \sigma_{21}) + \Omega_2(\sigma_{23} + \sigma_{32}). \tag{4}$$

Here $\Omega_1, \Omega_2$ are Rabi frequencies of acting fields and $\sigma_{kr} = |k\rangle\langle r|$, $k,r = 1,2,3$ are atomic projectors onto the subspaces generated by eigenvectors of Hamiltonian $H_0$ of pure atomic system:

$$H_0|k\rangle = E_k|k\rangle. \tag{5}$$

$\Delta_1 = \varepsilon_1$, $\Delta_2 = \varepsilon_2 - \varepsilon_1$ are single and two-photon detunings, $\varepsilon_1 = E_2 - E_1 - \omega_1$, $\varepsilon_2 = E_2 - E_3 - \omega_2$.

We will seek for the stationary density matrix of atom-field system which changes at the frequencies of acting modes and their integer combinations. We will use the following master equation in the stationary limit:

$$[H,\rho] + iG\rho = 0. \tag{6}$$

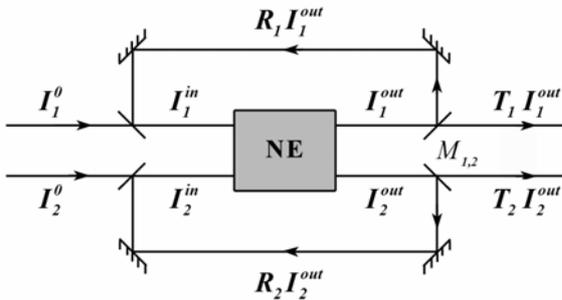
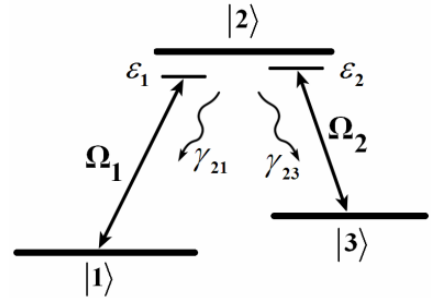

Figure 1. Unidirectional ring cavity with an atomic sample.   Figure 2. Three-state atom driven by electromagnetic fields.

Here $G$ is a relaxation superoperator:

$$(G\rho)_{ij} = \begin{cases} -\rho_{ij}\Gamma_{ij}, \ \Gamma_{ij} = \Gamma_{ji}, \ i \neq j; \\ \sum_{k=1}^{3}(\gamma_{kj}\rho_{kk} - \gamma_{jk}\rho_{jj}), \ i = j. \end{cases} \quad (7)$$

The absorption factors $\eta_j\left(I_1^{in}, I_2^{in}\right)$ may be evaluated in the following way:

$$\eta_j\left(I_1^{in}, I_2^{in}\right) = \exp\left(2k\,\text{Im}\sqrt{1+4\pi\chi_j}\right). \quad (8)$$

Here $k$ is wave number, $\chi_j$ is an electromagnetic susceptibilities at the transition $|j\rangle \to |j+1\rangle$, $j = 1,2$. It may be obtained as follows:

$$\chi_j = \frac{N_a |D_j|^2}{\hbar\varepsilon_0\Omega_j}\rho_{j,j+1}, \quad (9)$$

where $N_a$ is atomic concentration in the cell, $D_j$ is dipole moment of corresponding transition and $\rho_{j,j+1}$ is the density matrix element.

## RESULTS

### 3.1 Input intensity trajectories and stability criterion

Here we represent the solution of the nonlinear system (3) with coefficients of absorption from (8) which is obtained numerically. On fig.3a two curves representing the set of solutions for each equation of system (3) are shown. The input intensities $I_1^0$ and $I_2^0$ are fixed. The intersection points of these two curves correspond to solutions of the system. For the comparison these curves are depicted in the case then $R_1 = 0$, $R_2 = 0,6$ (fig.3b). It is evidently that switching on the second loop of feedback follows appearing additional solutions of (3). We will show further that in the domain $\left(I_1^0, I_2^0\right)$ there such solutions appear the hysteresis-type relation between input intensity of one field and output intensity of another field may be observed.

In analogous with system containing one loop of feedback there are stable and unstable solutions among all obtained from (3). An unstable solution is never appears at the output of the system, but transfer to the one of the stable solutions. In the case of one feedback loop an unstable solutions correspond to that peace of the input-output curve which has negative slope for the S-type systems and positive slope for the N-type systems. The case of two loops of feedback requires conditions which allow classifying all solutions gained. We will use the facts from bifurcation theory. Let us linearize system (3) and estimate the norm of the following matrix $L_{2\times 2}$:

$$L_{jk} = \frac{I_j^0 R_j}{\left[1+R_j\eta_j\left(I_1^{in}, I_2^{in}\right)\right]^2}\frac{\partial\eta_j\left(I_1^{in}, I_2^{in}\right)}{\partial I_k^{in}}, \ j,k = 1,2. \quad (10)$$

Stable solution of (3) corresponds to the point in which $\|L\|_2 < 1$. Required matrix norm may be obtained approximately as follows:

$$\|L\|_2 = \sqrt{\lambda_{\max}\left(L^T L\right)}. \quad (11)$$

Here $\lambda_{\max}(A)$ is a maximal eigenvalue of corresponding matrix $A$. Applying this criterion to the points depicted in fig.3a we can find that solutions 1 and 3 are stable and 2 is unstable. This method of determining stability is universal and may be used both in the case of one feedback loop and in the case then its number greater than one.

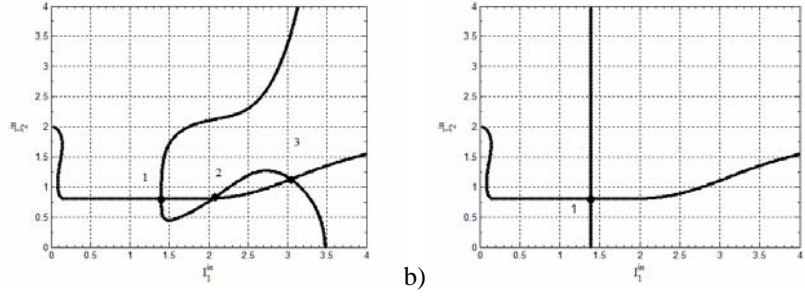

a)  b)

**Figure 3**. The solution of the nonlinear system (3) under certain quantity of input intensities $\left(I_1^0, I_2^0\right)$: a) two loops of feedback are switched on and $R_1 = R_2 = 0.6$; b) one loop of feedback is switched on ($R_1 = 0$, $R_2 = 0.6$); other parameters are represented in Appendix.

### 3.2 Bistability domain

Selecting the points $\left(I_1^0, I_2^0\right)$ where there are exist more than one solution $\left(I_1^{in}, I_2^{in}\right)$ of (3) one can obtain the set like are depicted on fig. 4 in the location of input intensities $I_1^0$, and $I_2^0$. This is an OB domain of absorptive type. In general it shares the plane on three different parts: interior domain $V_a$, OB domain and exterior domain $V_t$. It will be shown further that these parts correspond to the different absorptive roles in the atomic sample. For instance, in $V_a$ exponential absorption is observed (absorbing domain) while $V_t$ corresponds to linear absorption low (transparent domain). Bistability domain is a boundary region where the type of absorptive role changes. Its form and magnitude depends on the system parameters in a complicated manner. The main of these relations we consider qualitatively. In fig. 4a the bistability domain under different quantities of single photon detunings is depicted. It is expected that in the limit of two photon resonance $(\Delta_2 = 0)$ absorptive bistability domain disappears. Fig. 4b and 4c shows the dependence on the reflection coefficients of semitransparent mirrors $R_1$ and $R_2$.

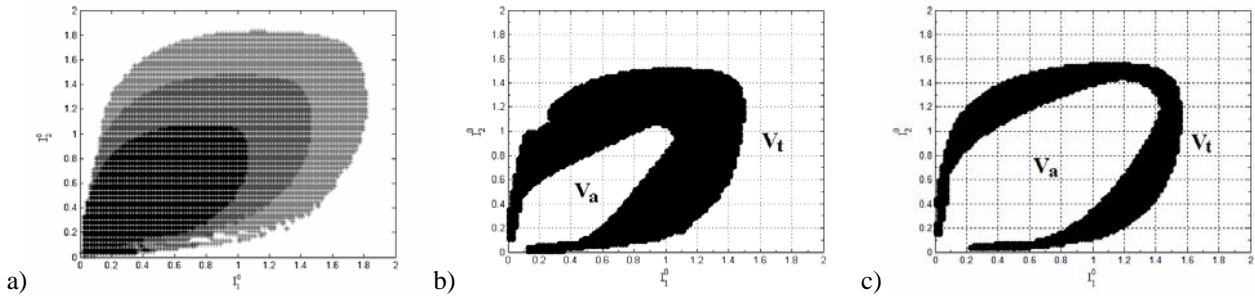

a)  b)  c)

**Figure 4**. OB domain: a) $R_1 = R_2 = 0.99$, $\varepsilon_1 = -\varepsilon_2 = 0.4$ (black); $\varepsilon_1 = -\varepsilon_2 = 0.7$ (dark grey); $\varepsilon_1 = -\varepsilon_2 = 1$ (grey); b) $R_1 = R_2 = 0.75$, $\varepsilon_1 = -\varepsilon_2 = 0.7$; c) $R_1 = R_2 = 0.6$, $\varepsilon_1 = -\varepsilon_2 = 0.7$; other parameters are represented in Appendix.

### 3.3 Input-output relations, cross-hysteresis

Here we will find the relations between input intensities $I_1^0$, $I_2^0$ and output intensities $I_1^{out}$, $I_2^{out}$ for different trajectories in the location of $I_1^0$, $I_2^0$. The parameters of OB domain are the same which used for fig. 4c to show. Let us start form the case then $I_2^0$ is fixed and $I_1^0$ varies. On figs. 5 the input – output dependences $I_1^{out}\left(I_1^0\right)$ and $I_2^{out}\left(I_1^0\right)$ for certain quantity of $I_2^0$ are depicted. There are hysteresis-type curves. The relation $I_2^{out}\left(I_1^0\right)$ shown on fig. 5b we called cross-hysteresis. As it was mentioned above the low level output signal corresponds to the quantities $\left(I_1^0, I_2^0\right)$, which lies in the domain $V_a$. The regions with high level output signal are conditioned by $\left(I_1^0, I_2^0\right)$ lying in the transparency domain $V_t$. Notice, the location of hysteresis branches which separate the regions of high and low output intensities are in correspondence with fig. 4c. Getting $I_1^0$ fixed and varying $I_2^0$ follows the similar results.

Modifying one of the input intensities provides the way to control the hysteresis and cross-hysteresis parameters. To advance the facilities of input-output controlling we can vary the input intensities in more complicate manner. For the instance they can be changed by obeying some role. Here we consider the particular case when whole OB domain is passed through. We will use parametrically coupled input intensities $I_1^{out}(t)$, $I_2^{out}(t)$ for it. The required trajectory is shown on fig. 6a. Here the points lying on the curve is the quantities of parameter $t$ which are used for convenience in constructing of further figures. The output intensities as functions of parameter $t$ corresponding to considered trajectory are depicted on fig. 6b and 6c. The forward and backward passes are marked by the arrows. Their essential difference is conditioned by different rates between initial quantities of input intensities. Namely, in the case of forward pass we have $I_1^0 \ll I_2^0$ and the second field makes corresponding atomic transition transparent while another transition leaves "dark". In the opposite case of backward pass we have $I_1^0 \gg I_2^0$ and atomic transition $|1\rangle \to |2\rangle$ is transparent while another is "dark". Choosing the point of leaving OB domain one can obtain a number of possibilities to get required thresholds and distance between different hysteresis branches. All of these essentially expand the facility of controlling the hysteresis.

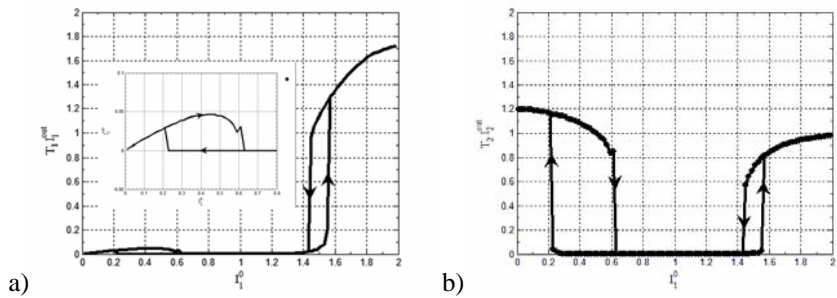

**Figure 5**. Input-output relations with $I_2^0$ fixed: a) hysteresis $I_1^{out}(I_1^0)$; b) cross-hysteresis $I_2^{out}(I_1^0)$. In both cases $R_1 = R_2 = 0.6$, $\varepsilon_1 = -\varepsilon_2 = 0.7$. Other parameters are represented in Appendix.

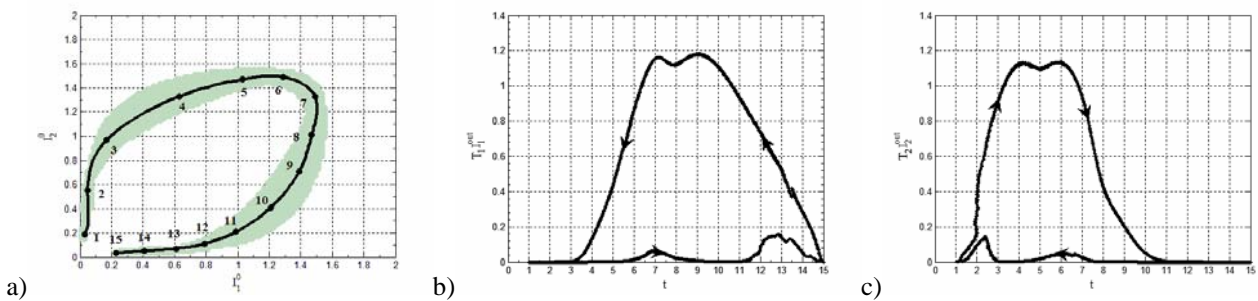

**Figure 6**. a) The trajectory passed throughout the OB domain; b) hysteresis-type curve $I_1^{out}(t)$; c) hysteresis-type curve $I_2^{out}(t)$. In both cases (b, c) $R_1 = R_2 = 0.6$, $\varepsilon_1 = -\varepsilon_2 = 0.7$. Other parameters are represented in Appendix.

### 3.4 Two-level scheme approximation

Now let us get the approximate analytical expressions for the quantities $\mathrm{Im}(\rho_{12})$ and $\mathrm{Im}(\rho_{32})$. We will use two-state atom approximation for it. Setting $\varepsilon_1 \approx \Omega_2 \gg \Omega_1$ and $\varepsilon_2 = 0$ one can get the following approximate expression for the imaginary part of $\rho_{12}$:

$$\mathrm{Im}(\rho_{12}) = \frac{\Omega_1 \gamma_{21} \Gamma_{21}}{4\Omega_1^2 \Gamma_{21} - \gamma_{21}\left((\Omega_2 - \Delta)^2 + \Gamma_{21}^2\right)}. \tag{12}$$

For nondiagonal density matrix elements we can find the following exact relation:

$$\frac{\mathrm{Im}(\rho_{21})}{\mathrm{Im}(\rho_{23})} = \frac{\gamma_{23}}{\gamma_{21}} \frac{\Omega_2}{\Omega_1}. \tag{13}$$

Substituting (13) into (12) we obtain:

$$\mathrm{Im}(\rho_{32}) = \frac{\Omega_1^2}{\Omega_2} \frac{\gamma_{21}^2}{\gamma_{23}} \frac{\Gamma_{21}}{4\Omega_1^2 \Gamma_{21} - \gamma_{21}\left((\Omega_2 - \Delta)^2 + \Gamma_{21}^2\right)}. \tag{14}$$

The OB domain gained due to analytical expressions (12) and (14) is depicted on figs. 7a. For comparison on fig. 7b the OB domain obtained from the exact counting with the same parameters is shown.

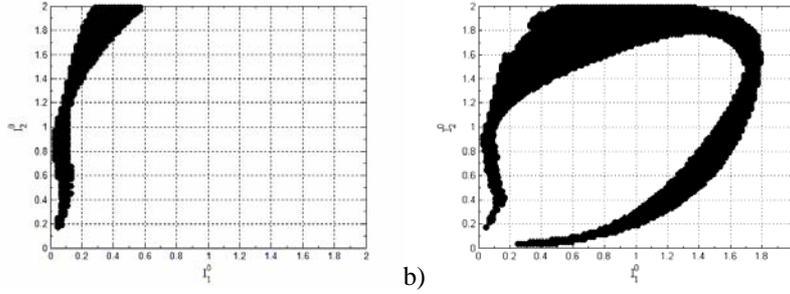

a)        b)

**Figure 7**. Comparison of analytical result in two-state atom approximation (a) and numerical counting (b). $R_1 = R_2 = 0.6$, $\varepsilon_1 = 2$. Other parameters are represented in Appendix.

## APPENDIX

Here we represent the systems parameters which were used for constructing the figures in the paper (in units of frequency $10^8 Hz$):

$$\gamma_{12} = \gamma_{32} = 3,$$
$$\Gamma_{12} = \Gamma_{21} = \Gamma_{23} = \Gamma_{32} = 0.5, \Gamma_{13} = 0, \tag{15}$$
$$L = 5cm, N_a = 10^{12} cm^{-3}, k = 0.5 \cdot 10^{-4} cm^{-1}, D = 10^{-18} CGSe.$$

## CONCLUSION

An absorptive type of optical bistability in the system surrounded by two loops of feedback is investigated in stationary limit. Solutions of nonlinear system of equations describing feedback are found and classified due to numerical counting. Two dimensional OB domain is obtained and dependence of its form and magnitude on the certain controlling system parameters is shown. The consideration of different trajectories passing through this domain allowed revealing the possibilities of hysteresis control. Particularly it allows observing the cross-hysteresis. The comparison of the results gained form numerical counting and analytical expressions in two-level system approximation are done.

Surely, more careful model must include transient process, the interference effects between modes of radiation and transfer effects in the cell. In spite of this imperfection the initial data which were taken from theoretical and experimental works [11, 13] and evaluated within the bounds of model presented here gives the results which are in agreement with results obtained in mentioned works.

## Acknowledgments

This study was supported by the Analytical Departmental Targeted Program "Development of Scientific Potential of Higher School (2009–2010), project 2.1.1/4215, by the Grant of the St. Petersburg Government for Students and Postgraduates of Higher Education Schools and Academic Institutes of St. Petersburg.

## References

[1] Arrathoon R., [Optical computing: digital and symbolic], Marcel Dekker inc., New York & Basel, 441 p. (1989).
[2] Wiswman H.M., "Quntum theory of continuous feedback", Phys. Rev. A, 49(3), 2133-2149 (1994).
[3] Kerckhoff J., Nurdin H. I., Pavlichin D. S., Mabuchi H., "Designing Quantum Memories with Embedded Control: Photonic Circuits for Autonomous Quantum Error Correction", Phys. Rev. Lett., 105(4), 040502-1/4 (2010).
[4] Lloyd S., "Coherent quantum feedback", Phys. Rev. A, 62(2), 022108-1/12 (2000).
[5] Rosanov N. N., [Optical bistability and hysteresis in distributive nonlinear systems], Nauka Fizmatlit, Moskow, 336 p. (1997). (in Russian)
[6] Mandel L., Wolf E., [Optical coherence and quantum optics], Cambridge University Press, Cambridge, 896 p. (1995).
[7] Agarwal G.P., Carmichael H.J., "Optical bistability through nonlinear dispersion and absorption", Phys. Rev. A, 19(5), 2074-2086 (1979).
[8] Bonifacio R., Lugiato L. A., "Optical bistability and cooperative effects in resonance fluorescence", Phys. Rev. A, 18(3), 1129-1144 (1978).
[9] Lugiato L. A., "Theory of optical bistability", Progress in Optics (edited by E. Wolf), 21, 69-216 (1984).
[10] Walls D.F., Zoller P., "A coherent nonlinear mechanism for optical bistability from three level atom", Optics Communications, 34 (2), 260-264 (1980).
[11] Harshawardhan W., Agarwal G. S., "Controlling optical bistability using electromagnetic-field-induced transparency and quantum interferences", Phys. Rev. A, 53(3), 1812-1817 (1996).
[12] Wang H., Goorskey D.J., Xiao M., "Bistability and instability in three level atoms inside an optical cavity", Phys. Rev. A, 65(1), 011801-1/4 (2001).
[13] Joshi A., Brown A., Wang H., Xiao M., "Controlling optical bistability in a three level atomic system", Phys. Rev. A, 67(4), 041801-1/4 (2003).
[14] Joshi A., Xiao M., "Optical multistability in three-level atoms inside an optical ring cavity", Phys. Rev. Lett., 91 (14), 143904-1/4 (2003).
[15] Agap'ev B.D., Gorniy M.B., Matisov B.G., Rozhdestvensky Yu.V., "Coherent population trapping in quantum systems", Uspekhi Fizicheskih Nauk, 163(9), 1-36 (1993). (in Russian)